\documentclass{article}
\usepackage{ijcai26}

\usepackage{times}
\usepackage{paralist}
\usepackage{microtype}
\usepackage{soul}
\usepackage{url}
\usepackage[hidelinks]{hyperref}
\usepackage[utf8]{inputenc}
\usepackage[small]{caption}
\usepackage{graphicx}
\usepackage{amsmath}
\usepackage{amsthm}
\usepackage{booktabs}
\usepackage{algorithm}
\usepackage[normalem]{ulem}
\usepackage{enumitem}

\usepackage{multirow}

\usepackage[switch]{lineno}
\usepackage{subcaption}
\usepackage{dsfont}
\usepackage{amsmath, amssymb, bm}
\usepackage{algorithm}
\usepackage{xcolor} 
\usepackage{caption} 
\usepackage{tabularx} 
\usepackage{makecell} 
\usepackage{optidef}
\usepackage[alignedforall]{lpform}
\usepackage{empheq} 
\usepackage{adjustbox}
\usepackage{enumitem}
\usepackage{algorithm}
\usepackage{algpseudocode}
\usepackage{amsmath}
\usepackage{amsthm}     % <-- add this
\definecolor{darkgreen}{RGB}{0,100,0}

\definecolor{lightgreen}{RGB}{37, 176, 30}
\definecolor{darkgreen1}{RGB}{35, 166, 27}
\definecolor{darkgreen2}{RGB}{20, 94, 16}

\definecolor{lightred}{RGB}{150,30,30}
\definecolor{darkred}{RGB}{255,30,30}
\definecolor{darkdarkred}{RGB}{155,0,0}

\newcommand\citet[1]{\citeauthor{#1}~[\citeyear{#1}]}

\usepackage{multirow}
\usepackage{algpseudocode}

\usepackage{xspace}
\usepackage{amsmath}

\DeclareRobustCommand{\algfmt}[1]{{#1}\xspace}
\DeclareRobustCommand{\algfmtnoitalic}[1]{{#1}\xspace}
\providecommand{\algThompsonstar}{\algfmt{TS*}}
\providecommand{\alggrdstar}{\algfmt{Greedy*}}

\providecommand{\smartmean}{\algfmt{SMaRT(E)}}
\providecommand{\smartmsmpl}{\algfmt{SMaRT(S)}}
\providecommand{\smart}{\algfmt{SMaRT}}
\providecommand{\smartGS}{\algfmt{SMaRT(GS)}}

\providecommand{\leastload}{\algfmt{Least Load}}
\providecommand{\upbnd}{\algfmt{Upper Bound}}
\providecommand{\overload}{\algfmtnoitalic{Overload-Case-Days/Med-Year}}
\providecommand{\overloadshort}{\algfmtnoitalic{OCDM}}
\title{SMaRT: Online Reusable Resource Assignment and an Application to Mediation in the Kenyan Judiciary}

\author{
Shafkat Farabi$^1$
\and
Didac Marti Pinto$^2$\and
Wei Lu$^3$\and
Manuel Ramos-Maqueda$^3$\and
Sanmay Das$^1$\and
Antoine Deeb$^3$\And
Anja Sautmann$^3$
\\
\affiliations
$^1$Virginia Tech\\
$^2$University College London\\
$^3$The World Bank\\
\emails{mfarabi@vt.edu, didac.marti.pinto.19@ucl.ac.uk, wlu4@worldbank.org, mramosmaqueda@worldbank.org, sanmay@vt.edu, adeeb1@worldbank.org, \\asautmann@worldbank.org}
}
\begin{document}
\nolinenumbers
\maketitle

\begin{abstract}
    Motivated by the problem of assigning mediators to cases in the Kenyan judicial system, we study an online resource allocation problem where incoming tasks (cases) must be immediately assigned to available, capacity-constrained resources (mediators). The resources differ in their quality, which may need to be learned. In addition, resources can only be assigned to a subset of tasks that overlaps to varying degrees with the subset of tasks other resources can be assigned to. The objective is to maximize task completion while satisfying soft capacity constraints across all the resources. The scale of the real-world problem poses substantial challenges, since there are over 2000 mediators, and a multitude of combinations of geographic locations (87) and case types (12) that each mediator is qualified to work on. Together, these features—unknown quality of new resources (newly onboarded mediators), soft capacity constraints (due to the mandate to assign cases without delay), and high-dimensional state space—make existing scheduling and resource allocation algorithms either inapplicable or inefficient. We formalize the problem in a tractable manner, using a quadratic program formulation for assignment and a multi-agent bandit style framework for learning. We demonstrate the key properties and advantages of our new algorithm, SMaRT (Selecting Mediators that are Right for the Task), compared with baselines on some stylized instances of the mediator allocation problem. We then turn to considering its application to real-world data on cases and mediators from the Kenyan Judiciary. SMaRT outperforms baselines and allows for controlling the tradeoff between the strictness of the capacity constraints and overall case resolution rates, both in situations where mediator quality is known beforehand and when the problem is bandit-like in that learning is part of the problem definition. On the strength of these results, we plan to conduct a randomized controlled trial in which we deploy SMaRT in the Judiciary's mediation management system. 
\end{abstract}

\section{Introduction}

As part of its Court Annexed Mediation (CAM) system, the Judiciary of Kenya diverts civil cases to court appointed mediators before they are tried in front of a judge. If a case fails to reach a settlement in mediation, it is escalated to a judicial trial. These mediators are accredited to operate in specific \emph{court stations} (serving different geographic regions) and for specific \emph{case types} (for example, family matters or business disputes). In addition to differences in which cases mediators are accredited to handle, there is also strong evidence that mediators differ in their ability to help parties reach a successful resolution. However, for reasons of time constraints, diminishing performance and equity, it is considered desirable to limit the number of cases each mediator is handling concurrently.

Once a case arrives into the system, a CAM officer is responsible for finding a suitable mediator who is accredited for the relevant court station and case type, and who is not over-committed in terms of the number of cases they are handling. The CAM officer communicates with the mediator, and, if the mediator is willing, assigns them the case.
Until recently, this process was  manual, which is both complicated and prone to inefficiencies. 
CAM officers must perform a complex, dynamic task that involves assessing mediator quality in addition to managing load across mediators. Moreover, any individual CAM officer may not account for the effects of their assignment decisions on overall mediation capacity, as mediators often work in more than one court station. Analysis of historical data and continual monitoring have established that manual assignment is inefficient (barely outperforming round-robin allocation in terms of resolution rates) and unequal (in terms of cases awarded across mediators conditioned on past performance), not to mention time consuming.\footnote{Source: Private communications.} 

For the reasons above, the Judiciary is interested in automating the process of case-mediator matching. If successful, an automated assignment tool can significantly  reduce the number of cases that are returned to the court for a judicial trial, increasing the successful resolution of cases between the parties and reducing the strain and backlog in the Judiciary. 

The first challenge in designing such a tool is to determine how good a mediator is at reaching case settlements. We quantify a given mediator's contributions to successful resolution by adapting value-added models, typically used to measure the performance of teachers, to our setting.  Continual entry and exit of mediators also means that the tool needs to balance exploring new mediators and exploiting identified better mediators, as in the Multi-Armed Bandit (MAB) literature \cite{MAL-068}. To accommodate uncertainty and learning, we maintain Gaussian belief distributions over mediator value added, centered around our point estimate.
MABs and their variants have, of course, been researched extensively and used in many real-world applications such as online recommendation systems, ad-auctions, etc. \cite{bouneffouf2020survey}. Mediator assignment resembles the Bandits-with-Knapsacks (BwK) model of \citet{BwK}, but with a few key differences. Notably, in the BwK model, there is a strict capacity limit for each resource over the entire course of time. In mediator assignment, this constraint is (1) dynamic, in that resources replenish -- when a mediator finishes with a case, their capacity is freed up -- and (2) soft, because mediators can temporarily go over capacity (and in manual assignment often do), although this is considered undesirable by the Judiciary. 
%Classical MAB algorithms, however, only strive to identify the ``best'' choice, and once identified with confidence, repeatedly choose it. For the mediator assignment problem, such MAB algorithms would not directly apply as the ``best'' choice could be at capacity. 

The second major challenge is that of online assignment across mediators with different accreditation sets (i.e., court stations and case types they are accredited for) and values added. The closest existing work that addresses a similar challenge is the OM-RR-PA online assignment of reusable resources algorithm \cite{dong2021efficient}. Again, there are several key differences with our setting. First, OM-RR-PA operates with hard capacity constraints on each reusable resource and allows delayed assignments of arriving tasks to avoid overload of resources. Second, it assumes that the time a resource will be in use once allocated is deterministic and known. Finally, OM-RR-PA does not address learning, and including the complexities of learning when mediators are accredited across many different possible subsets of tasks would be non-trivial in that setting. 
%First, OM-RR-PA  allows delayed assignment of arriving tasks and operates with hard capacity constraints on each reusable resource. 

In summary, mediator assignment requires a new approach, which is the goal of  this paper. We address all the challenges above with our proposed algorithm \smart (\textbf{S}electing \textbf{M}ediators that \textbf{a}re \textbf{R}ight for the \textbf{T}ask). \smart assignment decides about mediator assignments online. It uses econometric estimation of the value added of a mediator (the percentage point increase in the base probability of case resolution if that mediator is assigned to the case), supported by Bayesian posterior updates. A quadratic programming (QP) formulation penalizes over-capacity allocations. The relative weight given to case resolution rates vs.\ capacity constraint violations can be tweaked by tuning a parameter.

After validating the behavior of \smart  in two stylized simulations, we turn to our main empirical analysis, which uses real data on case arrivals and mediators in the  Kenyan Judiciary. We demonstrate the benefits of \smart relative to several sensible baselines, including round-robin allocation and capacitated versions of greedy  and Thompson Sampling. \smart typically outperforms baselines in maximizing agreement rates while maintaining reasonably low mediator overload rates. This validation provides justification for a pilot deployment of \smart in Kenya in the near future. 
%We carry out simulation on both synthetic and real-world data collected from the Kenyan Judiciary. We compare our proposed algorithm with well-established MAB algorithms such as greedy and Thompson-Sampling. We have found that our algorithm out-performs greedy and Thompson-Sampling in the simulations, and out-performs historic court-officer assignments in case completion rates as well as capacity management. 

\section{Background and Related Work}
% writing the explanation of why other consdered algorithms would not have worked for this problem:

% A summary of all the relevant other algorithms you considered, and why they won't work for this problem. We need to explain why our baselines are all simple rather than using some of the complex methodology that has been developed already for similar problems (bandits with knapsacks, my prior AAMAS paper on reusable resources and any others that have been developed since then, etc.)
%%%%%%%%%%%%%%%%%%%%%%%%%%%%%%%%%%%%%%%%%%%%%%%%%%%%%%%%%%%%%%%%%%%%%%%%
%\subsection{Judicial Efficiency and Mediation}

Prior work links judicial inefficiency to weaker economic performance \cite{kapopoulos2024judicial} and highlights substantial scope for improving court operations \cite{ferro2020benchmarking,ippoliti2020efficiency}. In Kenya, active case management has been associated with improved judicial efficiency and reduced crime rates \cite{odhiambo2014efficiency}. Nonetheless, civil cases remain severely backlogged. To address this, the Kenyan Judiciary introduced Court-Annexed Mediation (CAM) to reduce the time and cost of dispute resolution. Court trials often span years and incur substantial legal and administrative costs; mediation offers a faster, more flexible, and less adversarial alternative. Under CAM, civil cases are screened and referred to mediation, with court-appointed mediators facilitating negotiation between parties within a 70-day window. Cases that reach agreement are resolved; others proceed to trial. A key challenge to scaling CAM lies in mediator assignment: historical data indicate that mediators persistently achieving  low case agreement rates continue to receive cases, and many mediated disputes ultimately return to court. We therefore ask how an algorithmic approach that leverages case-level and mediator-level data can optimally assign mediators to incoming cases to improve overall case outcomes.

Multi-armed bandits (MAB) are a standard framework for exploration–exploitation problems where the value of different options is incompletely known \cite{MAL-068}. Bandits with Knapsacks (BwK) extend this framework to settings with resource constraints, where each arm pull yields a reward and incurs stochastic resource costs subject to global budgets \cite{BwK}. While conceptually related to our mediator load constraints, key differences outlined previously prevent direct application of BwK methods to our setting.
\citet{bernasconi2023bandits} extend BwK to Bandits with Replenishable Knapsacks, which handle deterministically replenishing resources but cannot accommodate the constraint violations that arise because cases must be assigned a mediator on arrival. \citet{concaveConvex} in turn generalize BwK by allowing penalties and diminishing returns, constraining the time-averaged outcome vector to lie in a given convex set and defining total reward as a concave, Lipschitz-continuous function of the outcome vector. However, their formulation does not support replenishable resources. 

MABs are widely used to learn agent quality in online resource allocation and multi-agent systems \cite{tran2012efficient,ho2016adaptive,liu2017sequential}.  MABs are employed for adaptive worker selection under budget and load constraints in crowdsourcing \cite{rangi2018multi} and for online matching of compute  to incoming jobs \cite{xu2020online}. MABs have also been deployed in socially impactful domains, including health information delivery in low-resource settings \cite{mate2020collapsing,mate2022field,dasgupta2025bayesian,verma2023restless}, maternal healthcare \cite{boehmer2025optimizing,wang2023scalable,mate2021risk}, public health adherence \cite{liang2025context}, environmental monitoring \cite{martin2020multiarmed}, corporate social responsibility \cite{ron2021corporate}, and disaster management \cite{liang2024multi}.

%MABs have been used extensively for learning agent quality in the context of online resource allocation and multi-agent systems more generally \cite{tran2012efficient,ho2016adaptive,liu2017sequential,wang2020strategic}.  \citet{rangi2018multi} use MABs to model adaptive worker selection under budget and load constraints in the context of crowd-sourcing.  \citet{xu2020online} design an efficient online MAB algorithm for matching compute clusters to incoming jobs. MABs have also been applied to solve problems with social impact. Examples include health information call delivery in low-resource communities \cite{mate2020collapsing,mate2022field,dasgupta2025bayesian,verma2023restless}, maternal healthcare \cite{boehmer2025optimizing,wang2023scalable,mate2021risk}, encouraging adherence in public health programs  \cite{liang2025context}, environmental monitoring \cite{martin2020multiarmed}, corporate social responsibility \cite{ron2021corporate}, and disaster management \cite{liang2024multi}. 

% \citet{OuNetworkedRestless} explore adaptation of MABs for mobile intervention  including health clinics and food pantries. In addition to these, several works apply MABs for social good in diverse fields such as 

The other thread of relevant literature is on modeling online bipartite matching problems with linear programs (LP) to design allocation algorithms. This general technique has been applied to several domains, including online ad auctions \cite{ho2013adaptive,devanur2009adwords},  crowd-sourced task assignment \cite{ho2012online}, and organ transplantation \cite{li2019incorporating}. These works typically assume that agents are impatient and resources disposable. \citet{dong2021efficient} adapt reusable resources with patient agents in OM-RR-PA. Similarly, we model mediator allocation to cases using a bipartite graph and apply a quadratic programming framework for optimization. Mediator assignment must deal with several novel issues that OM-RR-PA does not confront, including stochastic resolution times, and soft capacity constraints. %The key difference being mediators can be assigned to upto 3 cases before reaching capacity, while the reusable resources have a capacity of 1. 
%Additionally, our cases are \textit{impatient} agents, and to accommodate that we allow for constraint violations for mediator capacities when deemed useful, and incorporate penalties associated with these constraint violations in the optimization objective. 
%\citet{dickerson2021allocation} also consider online bipartite matching with impatient agents and reusable resources. Their approach accounts for stochastic replenishment time, but differs from our work in that they do not facilitate soft capacity constraint violations. Furthermore, they assume each resource can be used by one agent at the same time, whereas our application requires multiple uses simultaneously.
\citet{dickerson2021allocation} study online bipartite matching with impatient agents and reusable resources under stochastic replenishment. Their framework does not permit the soft capacity constraint violations allowed in our setting and assumes each resource serves at most one agent at a time, whereas our application requires simultaneous multiple uses. 
The full-information mediator allocation problem can be formulated as an MDP \cite{puterman2014markov}, but this does not accommodate online learning of mediator VAs and scales poorly to real world scenarios due to state-space explosion. Constraint programming approaches (e.g., RCPSP \cite{CAVALCANTE2013433}) can capture mediator capacity and licensing constraints, but cannot capture case resolution rate optimization and immediate case assignments.

\section{SMaRT Assignment}
\subsection{Court Annexed Mediation (CAM) in Kenya}

In the CAM process, a CAM officer assigns a mediator to each incoming case. Mediators are accredited only for specific case types and court stations, so assignments are drawn from the pool of mediators eligible for the case’s (type,station) pair. The CAM secretariat prefers that mediators handle no more than three concurrent cases.

Until recently, CAM officers navigated these constraints manually, supported by a digital mediation platform called `Cadaster' that was developed with the support of the World Bank and tracks mediator accreditation and cases as well as mediator assignments and past case outcomes. Our value-added estimates (see below) suggest that even mediators that performed consistently below average were often assigned in preference to higher-performing mediators, and the 3-case capacity constraint was frequently needlessly violated. CAM is now piloting a ``smart assignment'' feature within Cadaster that automatically proposes the most suitable mediator for each case for CAM officer approval. The algorithm developed here is proposed as the back end of that feature.

\subsection{Value Added Estimation}
\label{sec:VAestimate}
\begin{figure}[t]
    %\centering [width=1\textwidth, trim=left bottom right top, clip]
    \centering 
    \includegraphics[width=0.5\linewidth,trim=0.2cm 0.3cm 0.2cm 0.0cm, clip]{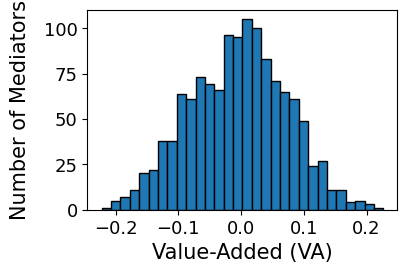}
    \caption{Histogram of empirical mediator VA estimates}
    \label{fig:medVAdistData}
\end{figure}

Inspired by the literature on the contribution of teachers to student learning outcomes, we model a mediator's contribution to case resolution using a value-added model \cite{kane2008estimating}. Analysis of historical data shows that mediator skill is idiosyncratic and not predicted by individual level-observables.\footnote{Deciding a mediator's case load \emph{a priori} only based on such observables would also violate ethical standards of non-discrimination.} To obtain a mediator's Value Added (VA), we estimate the probability of case resolution net of mediator contribution $p_i$ based on the regression $Y_{ij} = X_i \beta + \nu_j + \epsilon_{ij}$. $Y_{ij}=1$ if case $i$ assigned to mediator $j$ reaches agreement, and $0$ otherwise.  $X_{i}$ is a vector of controls including case type, court station, time period, and referral mode fixed effects, $\nu_j$ is a vector of mediator fixed effects, and $\epsilon_{ij}$ is the error term. We residualize outcomes as $\hat r_i = Y_{ij} - X_i\hat\beta$ and adapt  shrinkage to estimate VA for mediators with at least 2 cases. For each mediator $j$ with $n_j$ cases, we split cases chronologically into two groups and compute average residuals $\bar r_j^1$ and $\bar r_j^2$. We estimate case-level variance $\sigma_\epsilon^2$ using $\mathrm{Var}(\hat{r}_{i}-\bar{r}^{t}_{j})$, where $t=1,2$ indexes the (chronological) group,  mediator-level variance $\sigma_\mu^2$ via $\mathrm{Cov}(\bar r_j^1,\bar r_j^2)$, and group-level variance $\sigma_\theta^2$ as $\mathrm{Var}(\hat r_i) - \sigma_\mu^2 - \sigma_\epsilon^2$.
 Mediator VA is then estimated as $\hat\mu_j = \frac{\lambda_j }{n_j}\sum_{i=1}^{n_j} \hat r_i$, with shrinkage factor $\lambda_j = \frac{\sigma_\mu^2}{\sigma_\mu^2 + h_j}$ and $h_j = \frac{1}{2}\!\left(\sigma_\theta^2 + \frac{\sigma_\epsilon^2}{0.5 n_j}\right)$. From our estimate of $\sigma^2_{\mu}$, a one standard deviation increase in mediator VA corresponds to $13.36$ percentage points increase in the likelihood of case resolution. A histogram of mediator VAs estimated using this method from historical case allocations (2016--2025) is shown in Figure \ref{fig:medVAdistData}.

%We residualize $Y_{ij}$ by letting $\hat{r_i} = Y_{ij}-\hat{p_{i}}= Y_{ij} - X_i\hat{\beta}$ and then adapt the shrinkage procedure from \citet{kane2008estimating} to obtain an estimate of VA for mediators with at least two cases as follows:

%\textcolor{red}{Sanmay: Two things here. One is to inline all the equations unless we are explicitly using the numbering to reference them later in the text. The other is to add the histogram of VA that was in the poster. }

%\begin{enumerate}
%    \item Split a mediator's $n_j$ cases into two separate groups chronologically and average the residuals in each group to obtain $\bar{r_j^1}$ and $\bar{r_j^2}$
%    \item Estimate the case-level variance components $\sigma_{\epsilon}^2$ using $\text{Var}(\hat{r}_{it} - \hat{r}_i^t)$, where $t = 1,2$; the mediator-level variance component $\sigma_{\mu}^2$ by taking the covariance of two averages $\text{Cov}(\overline{r}^{1}_{j},\overline{r}^{2}_{j})$; and the group-level variance $\sigma_{\theta}^2$ component as the remainder $V(\hat{r_i}) - \sigma_{\mu}^2 - \sigma_\epsilon ^2$.
%    \item Estimate VA of mediator $j$ as $\hat{\mu}_j = \lambda_j \frac{1}{n_j}\sum_{i=1}^{n_j} \hat{r}_i$ with shrinkage factor $\lambda_j = \frac{\sigma_{\mu}^2}{\sigma_{\mu}^2+h_j}$, where $h_j = \frac{1}{2}[\sigma_\theta^2 + \frac{\sigma_\epsilon^2}{0.5n_j}]$.
%\end{enumerate}

\paragraph{Estimating Uncertainty and Making Efficient Updates}

%For many bandit algorithms like Thompson Sampling, we require not only a point estimate of VA, but also an estimate of the uncertainty in the point estimate. Moreover, estimating the VA with the method above is quite computationally expensive and cannot be performed each time a case is resolved. 

Many bandit algorithms, such as Thompson Sampling, require both point estimates of VA and associated uncertainty in those estimates to balance between exploration and exploitation of arms. Moreover, the above VA estimation procedure is computationally expensive and cannot be run after each case resolution. To resolve this, we define a Gaussian belief over $\mu_j$ centered around the VA point estimate $\hat{\mu}_j$. 
Recall that the  probability of successful resolution of case $i$ with base resolution probability $p_i$ assigned to mediator $j$ with VA $\mu_j$ is given by ($p_i+\mu_j$). We use an indicator random variable $Y_{ij}$ to denote whether the case is successfully resolved. We place a prior $\mu_j\sim\mathcal N(0,\sigma_\mu^2)$ and update after observing
$Y_{ij}$ via $f(\mu_j\mid Y_{ij}) \propto (p_i+\mu_j)^{Y_{ij}}(1-p_i-\mu_j)^{1-Y_{ij}}
\exp(-\mu_j^2/2\sigma_\mu^2)$. We further approximate the posterior with a Gaussian $\mathcal N(\mu_j \mid Y_{ij}, (\sigma_j \mid Y_{ij})^2)$ through moment matching. Upon observing an additional outcome $Y_{i'j}$, this posterior is used as the prior to update $f(\mu_j \mid Y_{ij}, Y_{i'j})$ sequentially. In practice, posterior means computed in this way initially deviate moderately from the explicit VA estimations above,\footnote{See Appendix 6.4 and 6.5 in the full version \cite{farabi2026smartonlinereusableresource}} but these deviations compound over multiple updates.  We therefore recalibrate the posteriors weekly by repeating the full VA estimation procedure and setting the new prior at $\mathcal{N}(\hat{\mu}_j,{(\sigma_{j}|Y_{ij},Y_{i'j},\dots})^2)$.

\subsection{Assignment as a Quadratic Program}
Our goal is to design an automated case allocation system that maximizes case resolution rate. Without capacity constraints, this would be a conventional MAB problem. Within each cell, the algorithm would simply identify the best arm (mediator), and then play that arm  repeatedly. However, local greedy assignment does not translate to global optimality in the presence of capacity constraints and overlapping accreditation sets as outlined above. There can be value in ``holding back'' a high-performing mediator based on the ``opportunity cost'' of not being able to assign this mediator to a future case. 

\begin{figure}[tb]
    
    %\centering [width=1\textwidth, trim=left bottom right top, clip]
    \centering
    \includegraphics[width=\linewidth,trim=5cm 1.8cm 7cm 2cm, clip]{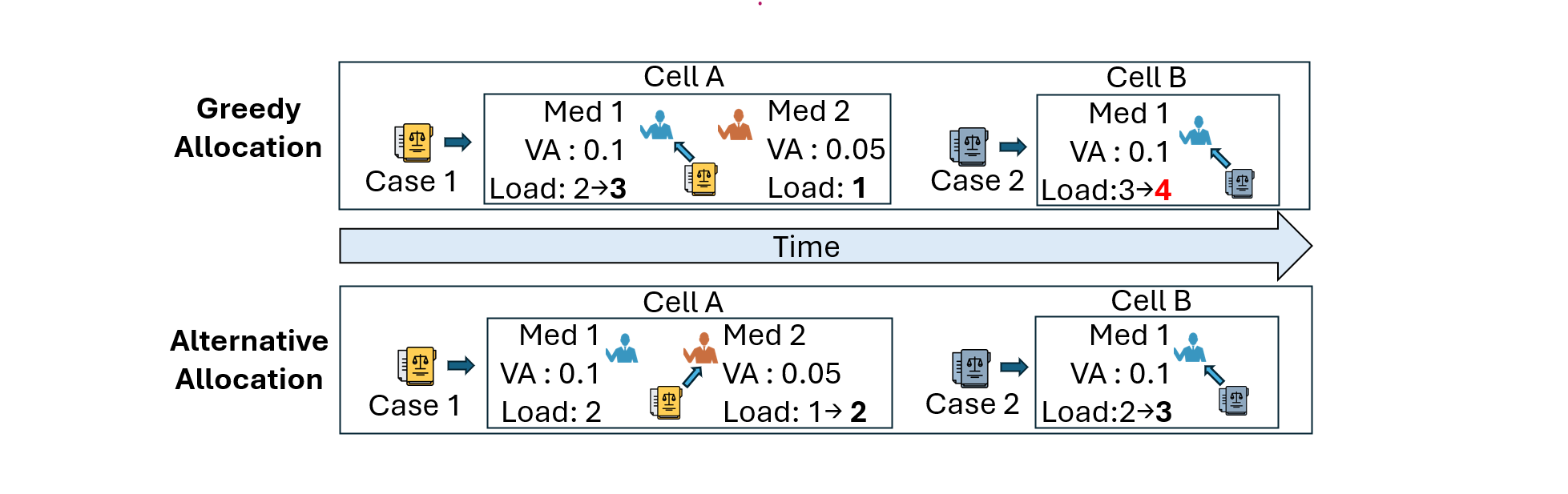}
    \caption{A hypothetical situation where holding back the best mediator is beneficial in the long run. Greedily assigning Case 1 in Cell A leads to overload when a new case arrives in Cell B.  Arrows indicate caseload changes after assignment. Red caseload indicates overload.}
    \label{fig:AllocationStrats}
\end{figure}
    
%Abstracting from learning and assuming perfect knowledge about mediator VAs, Figure \ref{fig:AllocationStrats} illustrates this with a stylized scenario. Consider mediators A and B, where $\mu_A > \mu_B$. There are two (case-type, court-station) cells. Both can work in Cell 1, but only A can  work in Cell 2. Mediator A already has two cases assigned to them, and mediator B has one case. Suppose one case arrives in Cell 1 and then another in Cell 2: under greedy allocation, mediator A will be allocated the first case, and will reach capacity. The case in Cell 2 will then put mediator A over capacity. A policy that holds back mediator A for the case in Cell 2 and allocates the case in Cell 1 to mediator B would achieve better capacity management, at the cost of a lower expected case resolution rate. By a similar logic, we can construct examples where holding a mediator back in one cell will prevent the assignment of a low performing mediator in another, thus raising overall case agreement rates.

Abstracting from learning, Figure~\ref{fig:AllocationStrats} illustrates a stylized scenario with  mediators 1 and 2, where $\mu_1>\mu_2$. Mediator~1 can serve both cells, while mediator~2 is restricted to Cell~A. If a case arrives in Cell~A followed by one in Cell~B, a greedy policy assigns the first case to mediator~1, leading to overload for this mediator when the next case arrives in Cell~B. Assigning the Cell~A case to mediator~2 instead preserves mediator~1 for Cell~B, improving capacity management at the cost of a lower expected agreement rate. By a similar logic, holding back high-performing mediators in one cell can prevent very low-performing assignments elsewhere, thereby increasing overall agreement rates.

Note that our algorithm must have  tolerance for case overloads. With stochastic arrivals and a requirement to immediately assign cases, there is a non-zero probability of all mediators in a cell reaching capacity. To find optimal allocations, we model a relaxed version of the problem as a mathematical program. Our approach accommodates caseload constraint violations by introducing a penalty term in the objective. This term is quadratic to deter the best mediator from successive overloads, making it a quadratic program (QP).

In the QP, mediators and cases are indexed by $u$ and $v$. $U$ denotes the set of all mediators. $V_r$ denotes the case(s) that have arrived prior to solving the QP. Future cases that may arrive are represented with shadow cases $V_{sh}$ sampled from a distribution. Mediator VA is $\mu_u$, and case fixed effects are $p_v$, interpreted as the probability of resolution with a zero-VA mediator. Mediator caseload and capacity are given by $L(u)$ and $C(u)$.
The allocation state is represented by a bipartite graph $G_s=(U\cup V,E)$, where $(u,v)\in E$ indicates that mediator $u$ is accredited for case $v$.
In the QP, $x_{uv}$ denotes the (possibly fractional) allocation of case $v$ to mediator $u$.
The model runs over a time horizon $T$ with timesteps $t$. $t_a(v)$ denotes arrival time for case $v$. We solve this QP with small time horizon (T=10) and assume case mediation durations outlast this short time horizon. 

\noindent
\begin{adjustbox}{max width=0.9\linewidth}
\begin{minipage}{1.1\linewidth}
\begin{alignat*}{3}
&\max_{x,\xi}\quad
&& \sum_{e\in E} x_e(\mu_u+p_v)\;-\;\lambda \sum_{u\in U}\xi_u^2
&&\\ 
&\text{s.t.}\quad
&& \sum_{e\in E(v)} x_e = 1 \tag{C1}\label{eq:c1}
&&  \forall\, v\in V_r \\[2pt]% \tag{C1}\label{eq:c1}\\[2pt]
&&& \sum_{e\in E(v)} x_e \le 1 \tag{C2}\label{eq:c2} 
&& \forall v \in V_{sh}\\[2pt]
&&& L(u)+\sum_{e\in E(u)} x_e \mathds{1}[t_a(v)\le t] \\
&&&\mkern100mu\ \le C(u)+\xi_u  
&&  \forall\, u\in U,\ \forall\, t\in [T] \tag{C3}\label{eq:c3} \\
%&&&\mkern100mu\ \le C(u)+\xi_u  
%&&  \forall\, u\in U,\ \forall\, t\in [T] \tag{C3}\label{eq:c3} \\ 
&&& 0 \le x_e \le 1
&&  \forall\, e\in E \tag{C4}\label{eq:c4} \\[2pt]
&&& \xi_u \geq 0
&&  \forall\, u\in U  \tag{C5}\label{eq:c5}
\end{alignat*}
\end{minipage}
\end{adjustbox}

\ref{eq:c1} enforces all real cases to be fully allocated, while \ref{eq:c2} allows shadow cases to go unassigned. \ref{eq:c3} imposes a soft capacity constraint ensuring mediators remain within capacity, with violations captured by the slack penalty term $\xi_u$. These violations are penalized quadratically in the objective to trade off expected case resolution against mediator overload. The parameter $\lambda$ controls the severity of overload penalty. %\ref{eq:c5} is added as a means to limit the solution space to make finding a solution more feasible. 
A discussion of the QP solvability and time complexity can be found in Appendix 6.1 in the full version \cite{farabi2026smartonlinereusableresource}.

\subsection{SMaRT Assignment Algorithm}

\begin{algorithm}[b]
\caption{SMaRT}
\label{alg:smart}
\footnotesize
\begin{algorithmic}[1]
    \State \textbf{Input:} $\lambda$, arriving case $v_r$, mediators $U$, history $H$, caseloads $L$, horizon $T$
    \State $\{\mu_u\}_{u\in U} \gets \mathrm{CalculateVA}(H)$
    \State $V \gets \{v_r\}\cup \mathrm{SampleCases}(0,T)$
    \State $E \gets \{(u,v): v\in V,\ u\in \mathrm{FindAccreditedMeds}(v)\}$
    \State $G_s \gets \mathrm{InitializeStateGraph}(V,E)$
    \State $QP \gets \mathrm{InitializeQP}(G_s,L,\lambda)$
    \State $x \gets \mathrm{Solve}(QP)$
    \State $\hat{u} \gets \arg\max_{u:(u,v_r)\in E} x_{u,v_r}$
    \State $\mathrm{Allocate}(v_r,\hat{u})$
    \State $H \gets H \cup \{(v_r,\hat{u})\}$; $L[\hat{u}] \gets L[\hat{u}] + 1$
\end{algorithmic}
\end{algorithm}

Now we bring everything  together in \smart (Algorithm \ref{alg:smart}). Records of unresolved cases ($L$) and mediation outcome history ($H$) for each mediator are maintained and updated as the algorithm executes. Upon case arrival, shadow cases are sampled and a state graph $(G_{s})$ is constructed with mediators, cases, and accreditation. The QP is then solved using Gurobi. This solution consists of fractional assignments of cases to mediators. Next, the algorithm looks at all the fractional assignments to mediators for the arriving case, and identifies the one ($\hat{u}$) with highest fractional assignment to the arriving case $v_r$.
 \smart allocates the case to this mediator. 

% This is updated whenever a new case is assigned and whenever an assigned case is resolved. We also maintain a history of all mediation outcomes in $H$. Whenever a case completes mediation, we record the outcome, mediator, and case specific factors in $H$ and update VAs.
%. Whenever a new case arrives, \smart calls on the method described in section 3.1 to calculate mediator VAs. 
%When a case arrives, the algorithm samples shadow cases from the case arrival distribution within a pre-determined future window. Next, accredited mediators are identified from the mediator pool for both the arriving and sampled shadow cases. Then, LP-Approximation is set up with the current state of the allocation problem. For this, the state graph ($G_{s})$ is constructed with mediators, cases, and accreditation. \textcolor{red}{The LP-Approximation is then solved using Gurobi. This solution consists of fractional assignments of cases to mediators. For the arriving case, we look at all the fractional assignments to mediators, and identify} the one ($\hat{u}$) with highest fractional assignment to the arriving case $v_r$.  \smart allocates the case to this mediator. 

\paragraph{Bandit Learning}

If the true mediator VAs are known, then the QP  reflects an approximation of the optimal assignment. Later we test this assumption in situations where we give \smart the true VAs (we refer to \smart with true VAs as \smartGS (for Gold Standard)). When the true VAs are unknown, we have to confront the exploration-exploitation dilemma. While some of this is mitigated in our setting by the mediator capacity constraints and the exploration they enforce, we also consider a variant of \smart that incentivizes exploration in a manner analogous to Thompson Sampling. The two versions of \smart we consider that incorporate learning differ only in how they implement \textit{CalculateVA(H)}:
\begin{itemize}[leftmargin=10pt,itemindent=0pt,labelsep=0.5em,noitemsep,topsep=0pt]
    \item \smartmean: As cases arrive, mediator VAs are defined to be the expectation (mean) of corresponding VA beliefs. 
    \item \smartmsmpl: As cases arrive, mediator VAs are sampled from corresponding VA beliefs, akin to Thompson sampling, encouraging exploration.
\end{itemize}

%In addition to this, for benchmarking purposes, we experiment with a version of SMaRT where true-VA information of the mediators are available to the algorithm. In this case no learning is necessary and we identify this version simply as \textit{SMaRT} in the results.

\section{Results}

\subsection{Benchmark Policies}
We compare \smart with the following allocation policies:
\begin{itemize}[leftmargin=10pt,itemindent=0pt,labelsep=0.5em,noitemsep,topsep=0pt]
    \item \leastload: The accredited mediator with lowest caseload is assigned to an arriving case. Ties are broken randomly.

    \item \alggrdstar: Assign each arriving case to the accredited mediator with the highest mean VA belief. Mediators at capacity are set aside until all other accredited mediators are equally overloaded.

    \item \algThompsonstar: Assign each arriving case to the accredited mediator with the highest sampled VA from the belief. Mediators at capacity are set aside until all other accredited mediators are equally overloaded.

    \item \upbnd: To upper-bound agreement rates, we ignore capacity and always assign the highest-VA accredited mediator by true VA.
\end{itemize}

We define \textbf{Agreement Rate}  as the share of arriving
cases that end in agreement rather than return to court and \textbf{\overload} (\overloadshort) as
$\frac{365}{mT}\sum_j\sum_d \max(OL_{j,d},0)$,
which expresses the degree of overload an average mediator experiences. Here $OL_{j,d}$ is mediator $j$’s case overload on day $d$. $T$ is the total runtime (in days), and $m$ is the number of mediators. Both metrics are averaged across multiple (128) independent runs when reported in the following sections.

%We define \textbf{\overload} (\overloadshort) as $\frac{\sum_j\sum_d \text{max}(L(j,d)-C(j),0)\times365}{m\times T}$ as an aggregate metric of mediator overloads. Here $L(j,d)$ is mediator $j$'s caseload on day $d$, $C(j)$ is the capacity constraint, $T$ is the total runtime in days, and $m$ is the number of mediators. This metric expresses the degree of overload an average mediator experiences.

%This metric expresses how many days per year the average mediator is overloaded (weighted by how much they were overloaded). 
%An alternative view of OCDM is the total number of days cases spent being classified as `overloaded' normalized per mediator per year. 

\subsection{Stylized Scenario Analysis}
To understand and validate the behavior of \smart, we crafted two small-scale scenarios with two cells and three mediators.
\begin{table}[tb]
\centering
\scriptsize
\resizebox{0.7\columnwidth}{!}{%
\begin{tabular}{crcc}
\toprule
\textbf{Mediator} & \multicolumn{1}{c}{\textbf{VA}} & \multicolumn{2}{c}{\textbf{Accreditations}} \\
\cmidrule(lr){3-4}
& & \textbf{Scenario 1} & \textbf{Scenario 2} \\
\midrule
Med-1 & 0.10  & Cell A, Cell B & Cell A, Cell B \\
Med-2 & 0.05  & Cell B         & Cell A \\
Med-3 & -0.10 & Cell B         & Cell B \\
\bottomrule
\end{tabular}
}
\caption{Mediator VAs and accreditations in two stylized scenarios.}
\label{tab:mediator-va-scenarios}
\end{table}

Mediator VAs and accreditations are shown in Table \ref{tab:mediator-va-scenarios}. Med-1 has the highest VA and is shared among cells A and B. Med-2 is slightly worse and depending on the scenario, is either restricted to cell A or B. Med-3 has the lowest VA and is restricted only to cell B. Cases arrive at a slightly higher rate at Cell A. The base $p$ values are set to $0.5$ for all cases.

%We assume that 0.055 cases arrive at cell A each day, and 0.05 cases at cell B. The base $p$ values are set to 0.5 for all cases. We simulate case allocation for the baselines and for \smart over a 365-day period 128 times and report average agreement rate of cases and \overloadshort. 
\begin{table}[tb]
\centering
\large
\resizebox{\columnwidth}{!}{%
\begin{tabular}{lcccccccc}
\toprule
\textbf{Algorithm} & $\lambda$ &
\multicolumn{2}{c}{\textbf{Known VA}} &
\multicolumn{2}{c}{\textbf{Learning + Mean VA}} &
\multicolumn{2}{c}{\textbf{Learning + Sampled VA}} \\
& &
\multicolumn{2}{c}{\textbf{(\smartGS)}} &
\multicolumn{2}{c}{\textbf{(\smartmean)}} &
\multicolumn{2}{c}{\textbf{(\smartmsmpl)}} \\
\cmidrule(lr){3-4}\cmidrule(lr){5-6}\cmidrule(lr){7-8}
& & \textbf{Agreement} & \textbf{\overloadshort}
  & \textbf{Agreement} & \textbf{\overloadshort}
  & \textbf{Agreement} & \textbf{\overloadshort} \\
& & \textbf{Rate} &
  & \textbf{Rate} &
  & \textbf{Rate} & \\
\midrule
\leastload & -- & 0.406 & 25.13 & -- & -- & -- & -- \\
\alggrdstar & -- & 0.439 & 41.71 & 0.425 & 36.31 & -- & -- \\
\algThompsonstar & -- & -- & -- & -- & -- & 0.404 & 34.17 \\
\midrule
\multirow{4}{*}{\smart}
& 0.01 & \textcolor{darkgreen}{0.453} & \textcolor{red}{94.12}
       & \textcolor{darkgreen}{0.438} & \textcolor{red}{79.16}
       & \textcolor{darkgreen}{0.437} & \textcolor{red}{67.67} \\
& 0.05 & \textcolor{darkgreen}{0.449} & \textcolor{red}{54.34}
       & \textcolor{darkgreen}{0.430} & \textcolor{red}{46.95}
       & \textcolor{darkgreen}{0.428} & \textcolor{red}{45.09} \\
& 0.10 & \textcolor{darkgreen}{0.444} & \textcolor{red}{44.85}
       & \textcolor{darkgreen}{0.427} & \textcolor{red}{39.03}
       & \textcolor{darkgreen}{0.427} & \textcolor{red}{38.46} \\
& 0.50 & \textcolor{red}{0.439} & \textcolor{darkgreen}{38.53}
       & \textcolor{red}{0.422} & \textcolor{darkgreen}{33.04}
       & \textcolor{darkgreen}{0.423} & \textcolor{darkgreen}{32.00} \\
\midrule
\upbnd & -- & 0.460 & 132.15 & -- & -- & -- & -- \\
\bottomrule
\end{tabular}
}
\caption{Scenario 1 results under different VA information assumptions. As expected, mediator overload (OCDM) decreases as $\lambda$ increases, but so does case agreement rate.}
\label{tab:scenario1}
\end{table}

\begin{figure}[tbh] % use [t] or [b] (bottom needs stfloats/dblfloatfix) left bottom right top
  \centering
  \includegraphics[width=0.7\columnwidth, trim=0in 0.05in 0in 0.05in, clip]{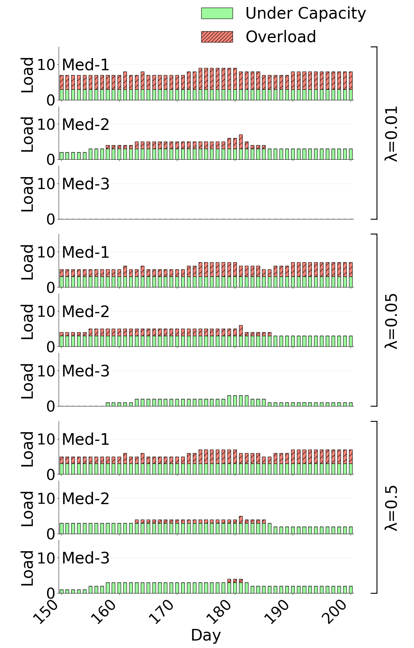}
  \caption{An example of case allocations (with known med VAs) in Scenario 1 by \smart. At lower $\lambda$, \smart favors overloading the best mediator (Med-1) to maximize case agreement rate.  At higher $\lambda$, \smart sacrifices agreement rate and focuses more on efficient case-load management, evidenced especially by the reduced red bars on Med-1 and the increased allocations  to Med-3.}
  \label{fig:case-distribution-smart}
\end{figure}

Table \ref{tab:scenario1} shows performance results in terms of agreement rates and average \overloadshort in Scenario 1. The ``Known VA'' column isolates the tradeoff between agreement rate and overload (we see similar patterns in the learning variants). \smart behaves as expected. With lower  $\lambda$s, it achieves better case agreement rates than \alggrdstar, at the cost of higher overload. As $\lambda$ increases, \smart distributes caseload more evenly, with some decline in case agreement rate. Figure \ref{fig:case-distribution-smart} visualizes this effect for \smartGS in individual runs of the simulation. At $\lambda=0.01$, Med-1 takes on the bulk of cases at both court-stations, leading to significant overloads (red) while Med-3 is never used. As the overload penalty $\lambda$ increases, \smartGS increases allocations to Med-2 and Med-3 to hold back  Med-1 for cases in Cell A. 

%Scenario 1 highlights the superior caseload management capabilities of SMaRT. Under `True VA' column of table \ref{tab:scenario1}, we present results when the algortihms have complete knowledge of mediator VAs. As there is no learning involved, we can isolate the caseload management capabilities SMaRT compared with the baselines. \textit{Greedy*} indiscriminately uses mediator 1 for cases arriving at both cells. Pretty soon mediator 1 reaches capacity and future cell B arrivals overload them. This is reflected in a higher \textit{overload case-day} metric in the table. SMaRT on the other-hand, can be tuned to overload mediator 1 to increase case resolutions, or to focus on better caseload management. To illustrate this in more detail, we present caseload by each mediator under different $\lambda$s in  Figure \ref{fig:case-distribution-smart}.  This leads to a lower resolution rate but also a lower overload case-days metric. 

%With learning enabled, resolution rates drop across the board for all algorithms in Table \ref{tab:scenario1}.\textit{Greedy*} and \textit{Thompson Sampling*} perform somewhat equivalently. Although, increased exploration by \textit{Thompson Sampling*} drives down mediator overloads. SMaRT-Mean and SMaRT-Sample perform equivalently at low $\lambda$s. Both of them achieve higher resolution rate with more overloads when compared with their baseline counterparts. SMaRT-Mean achieves slightly higher resolution rate while SMaRT-Sample achieves less overloads, likely due to more exploration. At higher $\lambda$s, SMaRT-Sample seems to have an edge over SMaRT-Mean, achieving both higher resolution rate and lower overloads. Sampling from VA belief pared with a stringent capacity constraint translates to increased exploration for SMaRT-Sample. This naturally means less overload. However, more exploration facilitates more informed mediator VA beliefs. This, in turn,drives up the resolution rate. 

% 2 Greedy: 0.4363 42.6718 0.4218 38.0286 

\begin{table}[tb]
\centering
\large
\resizebox{\columnwidth}{!}{%
\begin{tabular}{lcccccccc}
\toprule
\textbf{Algorithm} & $\lambda$ &
\multicolumn{2}{c}{\textbf{Known VA}} &
\multicolumn{2}{c}{\textbf{Learning + Mean VA}} &
\multicolumn{2}{c}{\textbf{Learning + Sampled VA}} \\
& &
\multicolumn{2}{c}{\textbf{(\smartGS)}} &
\multicolumn{2}{c}{\textbf{(\smartmean)}} &
\multicolumn{2}{c}{\textbf{(\smartmsmpl)}} \\
\cmidrule(lr){3-4}\cmidrule(lr){5-6}\cmidrule(lr){7-8}
& & \textbf{Agreement} & \textbf{\overloadshort}
  & \textbf{Agreement} & \textbf{\overloadshort}
  & \textbf{Agreement} & \textbf{\overloadshort} \\
& & \textbf{Rate} &
  & \textbf{Rate} &
  & \textbf{Rate} & \\
\midrule
\leastload & -- & 0.380 & 4.13 & -- & -- & -- & -- \\

\alggrdstar & -- & 0.406 & 5.18 & 0.394 & 5.06 & -- & -- \\

TS* & -- & -- & -- & -- & -- & 0.392 & 4.82 \\
\midrule
\multirow{4}{*}{\smart}
& 0.01 & \textcolor{darkgreen}{0.452} & \textcolor{red}{94.30}
       & \textcolor{darkgreen}{0.417} & \textcolor{red}{62.06}
       & \textcolor{darkgreen}{0.417} & \textcolor{red}{53.80} \\
& 0.05 & \textcolor{darkgreen}{0.440} & \textcolor{red}{43.90}
       & \textcolor{darkgreen}{0.403} & \textcolor{red}{25.34}
       & \textcolor{darkgreen}{0.407} & \textcolor{red}{29.29} \\
& 0.10 & \textcolor{darkgreen}{0.430} & \textcolor{red}{22.91}
       & \textcolor{darkgreen}{0.398} & \textcolor{red}{14.34}
       & \textcolor{darkgreen}{0.400} & \textcolor{red}{17.12} \\
& 0.50 & \textcolor{darkgreen}{0.410} & \textcolor{red}{5.43}
       & \textcolor{red}{0.394} & \textcolor{red}{5.54}
       & \textcolor{darkgreen}{0.394} & \textcolor{red}{5.67} \\
\midrule
\upbnd & -- & 0.460 & 132.15 & -- & -- & -- & -- \\
\bottomrule
\end{tabular}
}
\caption{Scenario 2 results under different VA information assumptions. Note the substantial gains in agreement rate at low $\lambda$s for SMaRT compared with \alggrdstar when VA is known. Similar to Scenario 1, Agreement Rate and OCDM  decrease with $\lambda$, as intended.}
\label{tab:scenario2}
\end{table}
%In Scenario 2, there is an alternative mediator available in both cells when Mediator 1 gets overloaded, but while cell A still has access to Mediator 2 with a net positive VA, cell B has to fall back on low-performing Mediator 3.

In scenario 2, both cells now have access to alternatives to Med-1. Cell A can fall back to Med-2 with a net positive VA, while cell B has to fall back to low-performing Med-3.
Holding Med-1 back in cell A can help avoid using Med-3 in cell B. We present simulation results from different settings in Table \ref{tab:scenario2}. \alggrdstar has a lower resolution rate but also less overload than in Scenario 1, because Med-3 now more often takes on cases in cell B. \smartGS with low $\lambda$ almost exclusively uses Med-1 as in Scenario 1. Even at higher $\lambda$s, \smartGS achieves higher resolution rates than \alggrdstar, at the expense of higher overloads, because \smartGS finds it beneficial to overload Med-1 slightly in cell B before assigning cases to Med-3. The dynamics are similar  with learning enabled.%but at high $\lambda$ there is no advantage over \alggrdstar.

\paragraph{Analysis of shadow prices}

\begin{figure}[b]
  \centering
  \includegraphics[width=\columnwidth, trim=0in 0cm 0in 0.1cm, clip]{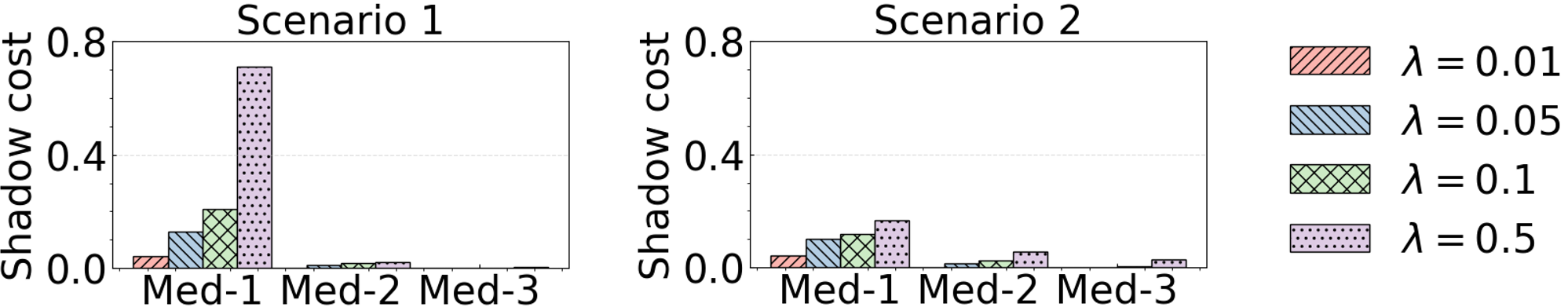}
 %left bottom right top
  \caption{Mean shadow price by mediator during allocation by SMaRT on the stylized examples for varying $\lambda$. As $\lambda$ increases, the shadow prices increase, demonstrating increased marginal benefit of holding them back for future cases. Note also Med-1's relatively higher value in Scenario 1, where only they can cover cases in Cell A, versus Scenario 2.}
  % old fig is Figures/duals.png
  \label{fig:duals}
\end{figure}

%Another sanity check on the behavior of the algorithm is to analyze mediator ``shadow prices'' which we obtain from the dual variables (Lagrange multipliers) associated with the QP constraints. This is also a useful exercise for eventually potentially converting the algorithm to a primal-dual method for online assignment (which is based on a mapping from mediator VAs and the characteristics of the state graph to these shadow prices). We  compute mean shadow prices during case allocation by \smart and report them in Figure \ref{fig:duals}. Shadow prices increase with $\lambda$ and are correlated with mediator VAs, as one would expect, but Mediator 1's shadow price is much higher than Mediator 2's, reflecting the value of Mediator 1's accreditation for Cell B. The effect is implicitly that cases in Cell 1 become relatively more likely to not be allocated to Mediator 1 -- the ``holding back'' mechanism. As $\lambda$ increases, the relative value of holding back also increases. The “shadow price” of a mediator reflects their value when assigned to a different case. The algorithm tries to assign the best possible (highest-value
%generating) mediator to a case, net of the value of ``holding them back” for a case where they would be even more valuable. Finally, how much the algorithm is willing to violate the capacity constraint is tunable though a single parameter that an administrator can control.

As a sanity check, we analyze mediator “shadow prices,” given by the dual variables of the QP constraints. These prices are also useful for interpretability of \smart as well as for future primal-dual online variants of the algorithm. Figure \ref{fig:duals} reports mean shadow prices under \smart. As expected, shadow prices increase with $\lambda$ and correlate with mediator VAs. However, Med-1’s shadow price is much higher than Med-2’s, reflecting Med-1’s added value for Cell B. This produces the intended ``holding back'' effect: cases in Cell B become less likely to use Med-1 when they may be more valuable elsewhere. As $\lambda$ increases, this effect strengthens. From an interpretability perspective, the ``shadow price'' of a mediator reflects their value when assigned to a different case. The algorithm tries to assign the best possible (highest-value
generating) mediator to a case, net of the value of ``holding them back” for a case where they would be even more valuable. Thus, \smart assigns mediators based not only on immediate VA but also on the opportunity cost of using them now, while allowing administrators to tune capacity violations through a single parameter.

%These results illustrate one of the strengths of \smart. An administrator can tune its focus along a spectrum. On extreme ends of this spectrum, \smart focuses only on case agreement or caseload management. This ``tunability'' gives \smart more flexibility than \alggrdstar or \leastload.

\subsection{Real Data}

\paragraph{Simulation framework}
We now turn to our main experiment on real data. We use mediation data collected from the Kenyan Judiciary covering the period 2016--2025. The dataset includes 30,633 cases involving roughly 2,100 mediators across 13 case types and 87 court stations, with information on arrivals, assignments, conclusion dates, outcomes, mediator accreditations, and station preferences. Based on this data, we develop a case assignment simulation to benchmark assignment strategies including \smartGS, \smartmean, \smartmsmpl and compare them. We use the Value Added model to compute the distribution of mediator VAs ($\mu$), as well as baseline probabilities of case resolution ($p$). Mediator VAs are  independently sampled from this distribution during simulations. We fit Poisson distributions for case arrival across (case-type, court-station) pairs on the data and sample case arrivals independently and identically. These distributions are also used to sample shadow cases within \smart. Case outcomes are sampled from 
$\mathrm{Bernoulli}\!\left(\Pi_{[0,1]}\!\left(p_{\text{case}}+\mu_{\text{med}}\right)\right)$. Survival analysis on the case mediation time indicates that duration is dependent on case-type and  outcome. We fit log-normal models to the mediation duration conditional on case-type and outcome for sampling mediation duration in the simulation.
For each setting, we average results over 128 seeded simulations. For a given seed, all inputs (case arrivals, mediator VAs, and accreditation patterns) are identical across methods. Since every arriving case must be assigned immediately, differences in performance reflect only how each method matches mediators to cases.

%For a simulated case of a given case type and the outcome sampled from the Bernoulli distribution above, we sample the conditional duration from the relevant fitted log-normal distribution. 

\paragraph{Results}

\begin{table}[tb]
\centering
\scriptsize
\resizebox{\columnwidth}{!}{%
\begin{tabular}{lcccc}
\toprule
\textbf{Algorithm} & $\lambda$ & \textbf{Agreement Rate} & \textbf{\overloadshort} & \textbf{Caseload Gini Index} \\
\midrule
\leastload & -- & 0.467 & 0.10 & 26.1\% \\

\alggrdstar & -- & 0.569 & 0.32 & 74.6\% \\
\midrule
\multirow{7}{*}{\smartGS}
& 0.01 & \textcolor{darkgreen}{0.617} & \textcolor{red}{52.68} & \textcolor{lightred}{87.3\%} \\
& 0.02 & \textcolor{darkgreen}{0.605} & \textcolor{red}{33.22} & \textcolor{lightred}{83.8\%} \\
& 0.03 & \textcolor{darkgreen}{0.596} & \textcolor{red}{24.30} & \textcolor{darkred}{81.8\%} \\
& 0.04 & \textcolor{darkgreen}{0.593} & \textcolor{red}{19.25} & \textcolor{darkred}{80.7\%} \\
& 0.05 & \textcolor{darkgreen}{0.590} & \textcolor{red}{15.96} & \textcolor{lightgreen}{79.8\%} \\
& 0.10 & \textcolor{darkgreen}{0.584} & \textcolor{red}{8.82} & \textcolor{lightgreen}{77.9\%} \\
& 0.50 & \textcolor{darkgreen}{0.574} & \textcolor{red}{0.94} & \textcolor{lightgreen}{74.7\%} \\
\midrule
\upbnd & -- & 0.659 & 122.01 & 96.5\% \\
\bottomrule
\end{tabular}
}
\caption{Simulations with true VA information available to the allocation algorithms on data collected from the judiciary. At $\lambda=0.01$, SMaRT achieves an agreement rate close to the upper bound but with much lower aggregate overload (OCDM) and much higher equity across mediators (Gini Index). As $\lambda$ increases, the agreement rate declines somewhat while overload and equity improve substantially.}
\label{tab:fulldataGRND}
\end{table}

Simulation results under full information of mediator VAs are reported in Table~\ref{tab:fulldataGRND}.
First, note that \smartGS with $\lambda=0.01$ attains a successful resolution rate that is close to the best achievable-- but with considerably better capacity management, since \upbnd by design gives cases only to the best mediator for each (court station, case type) cell. Relative to \alggrdstar, this is nearly a 8\% gain in performance, but it comes at the cost of substantially higher overload: OCDM rises to 52, compared to 0.32 under \alggrdstar. OCDM is an aggregate measure of our goal to minimize and equalize case overload. We therefore also compute Gini Index which shows the level of inequality in case assignment across mediators. With a high case influx rate, lower overloads correlate with equity of caseload distribution, which makes Gini Index a good comparative measure. As $\lambda$ increases, the agreement rate goes down but so do OCDM and the Gini Index, demonstrating the ``tunability'' of \smart.% in this setting.
%Large number of mediators available to take on cases translated to lower case overloads. Although, some overloading is inevitable as demonstrated by non-zero `overload case-days' incurred by `Lease Load' allocation. SMaRT with low $\lambda$ achieves a agreement-rate within 2p.p. `Upper-Bound'. The `overload case-days' incurred by SMaRT is only 15. This means each mediator on average spends 15 case-days being overloaded every year. While this metric is relatively high when compared with \textit{Greedy*}, we think the performance gain justifies spending about 2 case-weeks overloaded every-year. As $\lambda$ increases, agreement rate falls and caseload management improves. At $\lambda=0.5$, agreement rate is slightly below \textit{Greedy*} with less overload case-days. 

We now benchmark performance when mediator VAs are unknown and must be learned from observed case outcomes. These results are presented in Table \ref{tab:realDataLearn}. We consider two different initializations of the learning process for the mediators. In one, we initialize all mediator beliefs as $\mathcal{N}(0,\sigma_{\mu}^{2})$ -- this is the ``blank slate'' version where we know nothing about the mediators in advance, which could be used in an environment where the idea of VA estimation is entirely new, or where all the mediators arrive at the same time. In the second, VA beliefs are instead initialized in a manner that replicates what we might expect to be able to do when we deploy this system in practice for the Kenyan Judiciary. %At the beginning of each simulation, we sample mediator true VAs from $\mathcal{N}(0,\sigma_{\text{VA}}^2)$. Next, 
We iterate through case records from the prior six years and sample each case outcome using its $p$-value and the assigned mediator's VA. Using these outcomes, we compute an initial belief for each mediator VA. It is interesting to note that there is considerable variation in the number of cases mediators have mediated in the real data. As a result, the initial beliefs exhibit substantial variation in their uncertainty. Some highly experienced mediators have narrow beliefs concentrated near their true VA, whereas newer mediators have wider beliefs centered near $0$.

First, consider the results from the blank slate experiment on the left panel of Table \ref{tab:realDataLearn}. \alggrdstar attains a higher agreement rate than \algThompsonstar, while the latter is  more equitable and leads to less OCDM, a byproduct of greater exploration. Both \smartmean and \smartmsmpl behave similarly to how \smartGS behaves with known VAs: they deliver higher agreement rates at smaller $\lambda$s, and increasing equity and reduced \overloadshort as $\lambda$ increases. At small  $\lambda$s they outperform their relevant baselines (\alggrdstar for \smartmean and \algThompsonstar for \smartmsmpl) in terms of case agreement rates. In this experiment, \algThompsonstar (and \smartmsmpl) are strictly worse than \alggrdstar (and \smartmean). This is because capacity management itself serves to induce sufficient exploration. It is notable that \smartmean consistently exceeds \alggrdstar in agreement rate. %and at $\lambda=0.5$, achieves slightly better performance on both agreement and caseload management.

%The incremental gains in agreement and the deterioration in caseload management for \textit{SMaRT-Sample} are more muted than for \textit{SMaRT-Mean}. Moreover, 

%While these simulations give us a good idea of how the algorithms perform when starting with no information regarding mediator VAs, they are also unrealistic. An implementation of this algorithm in the real world will have access to years worth of data collected on case allocations. To understand the algorithms' behavior under such circumstances,  Results from these simulations are also included in Table \ref{tab:realDataLearn}. 
\begin{table}[t]
\centering
\large
\resizebox{\columnwidth}{!}{%
\begin{tabular}{lcccccccc}
\toprule
\textbf{Algorithm} & $\lambda$ &
\multicolumn{3}{c}{\textbf{Initial belief set to $\mathcal{N}(0,\sigma_{\mu}^2)$}} &
\multicolumn{3}{c}{\textbf{Initial belief calculated from data}} \\
\cmidrule(lr){3-5}\cmidrule(lr){6-8}
& & \textbf{Agreement} & \textbf{\overloadshort} & \textbf{Caseload}
  & \textbf{Agreement} & \textbf{\overloadshort} & \textbf{Caseload} \\
& & \textbf{Rate} & & \textbf{Gini Index}
  & \textbf{Rate} & & \textbf{Gini Index} \\
\midrule
\alggrdstar & -- & 0.489 & 0.44 & 68.4\% & 0.473 & 0.38 & 63.1\% \\
\midrule
\multirow{4}{*}{\smartmean}
& 0.01 & \textcolor{darkgreen}{0.518} & \textcolor{red}{72.17} & \textcolor{lightred}{86.4\%}
       & \textcolor{red}{0.448} & \textcolor{red}{37.22} & \textcolor{lightred}{74.0\%} \\
& 0.05 & \textcolor{darkgreen}{0.499} & \textcolor{red}{24.69} & \textcolor{darkred}{76.1\%}
       & \textcolor{red}{0.461} & \textcolor{red}{8.77} & \textcolor{darkred}{64.6\%} \\
& 0.10 & \textcolor{darkgreen}{0.494} & \textcolor{red}{14.77} & \textcolor{darkred}{73.2\%}
       & \textcolor{red}{0.462} & \textcolor{red}{4.25} & \textcolor{darkred}{62.9\%} \\
& 0.50 & \textcolor{darkgreen}{0.491} & \textcolor{red}{4.20} & \textcolor{darkred}{69.6\%}
       & \textcolor{red}{0.465} & \textcolor{red}{0.57} & \textcolor{lightgreen}{61.0\%} \\
\midrule
\algThompsonstar & -- & 0.484 & 0.34 & 53.7\% & 0.489 & 0.31 & 50.8\% \\
\midrule
\multirow{4}{*}{\smartmsmpl}
& 0.01 & \textcolor{darkgreen}{0.499} & \textcolor{red}{27.31} & \textcolor{darkdarkred}{61.0\%}
       & \textcolor{darkgreen}{0.495} & \textcolor{red}{9.31} & \textcolor{darkred}{52.3\%} \\
& 0.05 & \textcolor{darkgreen}{0.488} & \textcolor{red}{8.08} & \textcolor{darkgreen2}{53.1\%}
       & \textcolor{darkgreen}{0.492} & \textcolor{red}{3.17} & \textcolor{lightgreen}{49.1\%} \\
& 0.10 & \textcolor{darkgreen}{0.488} & \textcolor{red}{4.31} & \textcolor{darkgreen2}{51.4\%}
       & \textcolor{darkgreen}{0.490} & \textcolor{red}{1.48} & \textcolor{darkgreen}{47.3\%} \\
& 0.50 & \textcolor{red}{0.484} & \textcolor{red}{0.54} & \textcolor{darkgreen2}{49.5\%}
       & \textcolor{red}{0.489} & \textcolor{darkgreen}{0.29} & \textcolor{darkgreen}{47.3\%} \\
\bottomrule
\end{tabular}
}
\caption{Simulations on data collected from the judiciary where VAs need to be learned during allocation. SMaRT(E) works well from a ``blank slate'' initialization, where all mediator VAs must be learned, since the capacity constraint leads to sufficient exploration. SMaRT(S) performs much better when VAs are initialized from real data, because it can better explore mediators on whom the algorithm has collected little data.}
\label{tab:realDataLearn}
\end{table}

When we turn to the ``warm start'' paradigm where initial beliefs are calculated from data (the right panel of Table \ref{tab:realDataLearn}), this picture changes. Now, \algThompsonstar performs significantly better than \alggrdstar both in agreement rate and in managing overloads. The explanation for this lies in uneven knowledge regarding mediator VAs. \alggrdstar focuses on exploiting experienced mediators it already has accurate estimates of VA for, neglecting to explore less experienced mediators, and thus failing to identify inexperienced mediators with high VA. Conversely, \algThompsonstar does explore these mediators and is able to exploit the high true VA mediators among them. 

For the same reason, but exacerbated because of its primary focus on high agreement rates and more willingness to tolerate overload, with low $\lambda$, the behavior of \smartmean is much poorer than in other experiments (and worse than \alggrdstar). Increasing $\lambda$ in this case leads to benefits in not just equity across mediators but also in agreement rates by inducing exploration as a side effect. \smartmsmpl is by far the best performer, with the sampling inducing exploration even at low $\lambda$, and higher values of $\lambda$ inducing more equity in allocation across mediators without significant reduction in case agreement performance in this case.

\section{Conclusion}
%\textcolor{red}{This project stems from a six-year collaboration between the World Bank  and the Kenyan Judiciary to modernize CAM through Cadaster, a digital mediation platform developed with substantial World Bank technical support to digitize workflows and identify bottlenecks in manual assignment. Cadaster’s “smart assignment” feature recommends mediators for CAM officer approval, with SMaRT queued as its back end. The design is grounded in Judiciary operational needs, with the authors advising the IT team and regularly coordinating with court officers. SMaRT has been approved for a year-long RCT with ethics approval; if it significantly improves agreement rates and mediator overload, the control group will be transitioned to SMaRT in consultation with the Judiciary.}

This paper stems from a six-year collaboration between the World Bank  and the Kenyan Judiciary to modernize CAM through Cadaster, a digital mediation platform developed with substantial World Bank technical support to digitize workflows and identify bottlenecks in manual assignment. The paper describes the design of an algorithm --SMaRT-- to confront a real challenge faced by the Kenyan Judiciary. SMaRT's design is grounded in Judiciary operational needs, with the authors advising the IT team and regularly coordinating with the CAM Secretariat and communicating with court officers. While in theory the problem could map to a number of different algorithmic approaches in the literature (bandits, online matching for resource allocation, etc.), when we worked with stakeholders  we discovered nuances that made direct application of existing approaches infeasible. In particular, both in practice and conceptually -- since case assignment cannot be delayed -- the capacity constraints on mediators are \emph{de facto} soft constraints. This feature combined with the accreditation restrictions and the complex nature of learning when new mediators can enter the system raise a number of interesting challenges that necessitated the development of a new algorithm. We believe that there is a range of interesting theoretical and algorithmic work to be done in this space, as  governments face similar allocation problems in non-market settings in many domains. SMaRT is queued to be Cadaster's “smart assignment” feature backend for automated mediator recommendation to CAM officers. SMaRT has been approved for a year-long RCT with ethics approval; if it significantly improves agreement rates and mediator overload, the control group will be transitioned to SMaRT in consultation with the Judiciary.

%The next steps for our project involve a pilot deployment and evaluation of SMaRT in collaboration with the Kenyan judiciary.

\section*{Ethical Statement}

This project is conducted in close collaboration with the Kenyan Judiciary and is designed to reflect the operational requirements of the governing agency. This study was approved by the National Commission For Science, Technology \& Innovation of Kenya (NACOSTI) and the ethics review committee at Kenyatta University, Nairobi. The data used in our experiments were collected by the Kenyan Judiciary and provided to the authors in de-identified form, with all personal information about mediators and case parties removed to protect privacy. The de-identified data will be made publicly available in the World Bank microdata library (https://microdata.worldbank.org) and the code base will be available through the World Bank's reproducible research repository (https://reproducibility.worldbank.org/) after it has gone through mandatory external reproducibility verification. 

\section*{Acknowledgments}
This work is partially supported by NSF Award 2533162, the World Bank's Knowledge for Change Program, and by the World Bank's Global Program on GovTech and Public Sector Innovation through the GTGP Trust Fund.
The authors thank our project partners, Hon.~Caroline Kendagor, Hon. Grace Sitati, Hon. Moses Wanjala, Clifford Ogutu, Court-Annexed Mediation (CAM), and the Mediation Accreditation Committee (MAC).

\bibliographystyle{named}
\bibliography{ijcai26}

\clearpage

\section{Appendix}

\subsection{Solvability of the QP}

We show that the quadratic program in Section~3.3 is a convex quadratic program and that, under the natural assumption that every real case has at least one eligible mediator, it admits an optimal solution.

Let $z=(x,\xi)\in \mathbb{R}^{|E|+|U|}$, where $x$ denotes assignment variables and $\xi$ denotes mediator-level overload slack variables. Let $c_e=\mu_u+p_v$ for $e=(u,v)$. The QP can be written as
\[
\max_{x,\xi}\; c^\top x-\lambda \sum_{u\in U}\xi_u^2,
\]
subject to constraints (C1)--(C5). Equivalently, multiplying the objective by $-1$, this is the minimization problem
\[
\min_{x,\xi}\; -c^\top x+\lambda \sum_{u\in U}\xi_u^2 .
\]
The Hessian of this minimization objective with respect to $z=(x,\xi)$ is block diagonal:
\[
Q =
\begin{bmatrix}
0 & 0\\
0 & 2\lambda I_{|U|}
\end{bmatrix}.
\]
For any conformable vector $w=(w_x,w_\xi)$,
\[
w^\top Qw = 2\lambda \|w_\xi\|_2^2 \ge 0
\]
whenever $\lambda\ge 0$. Thus $Q\succeq 0$, and the minimization form has a convex quadratic objective. Since constraints (C1)--(C5) are all linear equalities or inequalities, the QP is a convex quadratic program.

It remains to establish feasibility and existence of an optimum. Define the assignment polytope

\[
P=
\left\{
x:
\begin{array}{l}
\sum_{e\in E(v)}x_e=1 \quad \forall v\in V_r,\\[2pt]
\sum_{e\in E(v)}x_e\le 1 \quad \forall v\in V_{sh},\\[2pt]
0\le x_e\le 1 \quad \forall e\in E
\end{array}
\right\}.
\]

Assume that every real case $v\in V_r$ has at least one eligible mediator, i.e., $E(v)\neq \emptyset$. Then $P\neq\emptyset$: for each real case $v$, choose one incident edge $e_v\in E(v)$ and set $x_{e_v}=1$; set all other real-case incident variables and all shadow-case variables to zero. This satisfies (C1), (C2), and (C4). Since $P$ is defined by finitely many linear constraints and box constraints, it is compact.

For any fixed $x\in P$, define the smallest nonnegative slack satisfying the capacity constraints as

\[
\resizebox{\columnwidth}{!}{$
\begin{aligned}
\xi_u^*(x)
&=
\left[
\max_{t\in[T]}
\left\{
L(u)
+
\sum_{e=(u,v)\in E(u)}
x_e\mathbf{1}[t_a(v)\le t]
-
C(u)
\right\}
\right]_+ .
\end{aligned}
$}
\]

where $[a]_+=\max\{a,0\}$. Then $(x,\xi^*(x))$ satisfies (C3) and (C5). Hence the full feasible set is nonempty.

Moreover, for any fixed $x\in P$, choosing any $\xi_u>\xi_u^*(x)$ cannot improve the objective when $\lambda>0$, because the objective subtracts $\lambda \xi_u^2$. When $\lambda=0$, the objective is independent of $\xi$, so $\xi^*(x)$ may still be chosen without loss of optimality. Therefore, the QP is equivalent, for purposes of attaining an optimum, to maximizing the continuous function
\[
\phi(x)
=
c^\top x
-
\lambda \sum_{u\in U}
\left(\xi_u^*(x)\right)^2
\]
over the compact set $P$.

The function $\xi_u^*(x)$ is continuous because it is the positive part of the maximum of finitely many affine functions of $x$. Therefore $\phi(x)$ is continuous. By the Weierstrass extreme value theorem, $\phi$ attains a maximum on $P$. Let $x^\star$ be a maximizer and set $\xi^\star=\xi^*(x^\star)$. Then $(x^\star,\xi^\star)$ is feasible for the original QP and attains the optimal objective value. Thus the QP admits at least one optimal solution.

\subsection{Runtime Analysis of SMaRT:}
In SMaRT, each arriving case is solved independently, so the runtime of a single QP depends on the number of sampled shadow cases and the number of eligible mediators for that arriving case, rather than on the total duration of the simulation. Since the QP is convex with positive semi definite $Q$, there exist algorithms that can find solutions in polynomial time \cite{ye1989extension}, \cite{polynomial}. For example, standard interior-point analysis for convex QPs, with $n = |E| + |U|$ variables and $m = O\!\left(|V_r| + |V_{sh}| + |U|T + |E| + |U|\right)$ constraints, the worst-case per-iteration complexity is $O\!\left((n+m)^3\right)$. In practice, we solve the QP using Gurobi, which exploits sparsity and optimized barrier-based routines for continuous convex QPs, and we observe substantially faster empirical runtime than this generic worst-case bound. The total runtime of SMaRT scales this polynomial complexity of solving each QP linearly in the number of cases assigned in total.

In our full-scale experiments presented in Tables \ref{tab:fulldataGRND} and \ref{tab:realDataLearn}, the median instance involved 52 total cases and 701 eligible mediators, yielding a median QP size of 3,632 variables and 14,326 constraints. The median solution time was 0.25 seconds, with the 95th and 99th percentiles at 1.23 and 2.68 seconds, respectively.

\subsection{Linear Approximation of the QP}
The QP in section 3.3 can be approximated using an LP in a small local region in the following manner:

\textbf{LP-Approximation}

\noindent
\begin{adjustbox}{max width=0.9\linewidth}
\begin{minipage}{1.03\linewidth}
\setlength{\jot}{5pt}
\begin{alignat*}{3}
&\max_{x,\xi}\quad
&& \sum_{e\in E} x_e(\mu_u+p_v)\;-\;\lambda \sum_{u\in U}\xi_u\times L(u)
&&\\ \\[1pt]
&\text{s.t.}\quad
&& \sum_{e\in E(v)} x_e = 1 \tag{C6}\label{eq:c51}
&&  \forall\, v\in V_r \\[2pt]% \tag{C1}\label{eq:c1}\\[2pt]
&&& \sum_{e\in E(v)} x_e \le 1 \tag{C7}\label{eq:c6} 
&& \forall v \in V_{sh}\\[2pt]
&&& L(u)+\sum_{e\in E(u)} x_e \mathds{1}[t_a(v)\le t] \\
&&&\mkern100mu\ \le C(u)+\xi_u  
&&  \forall\, u\in U,\ \forall\, t\in [T] \tag{C8}\label{eq:c7} \\ 
&&& 0 \le x_e \le 1
&&  \forall\, e\in E \tag{C9}\label{eq:c8} \\[2pt]
&&&0\le \xi_u \le \max(L(u) - C(u)+1,0)
&&  \forall\, u\in U  \tag{C10}\label{eq:c9}
\end{alignat*}
\end{minipage}
\end{adjustbox}

Note the replacement of the quadratic term $\lambda \sum_{u \in U} \xi_u^2$ in the objective with $\lambda \sum_{u \in U} \xi_u \times L(u)$. Here $L(u)$ is the current case-load of mediator $u$ at the time a new case arrives. For an optimal QP solution, $\xi_u=\max(0,L'(u)-C(u))$, where $L'(u)$ is the mediator case load in that optimal solution. It follows that $\sum_{u \in U} \xi_u^2 =  \sum_{u \in U} \xi_u \times (L'(u)-C(u)) \approx \sum_{u \in U} \xi_u \times L'(u)$. If $L(u)$ is the caseload during the LP formulation, and if we enforce $\xi_u\leq L(u)-C(u)+1$ in the LP, then we could approximate the penalty term as $\lambda \sum_{u\in U} L(u)$. This approximation is good only for values of $\xi_u$ that are near $L(u)$. This could be useful when solving a QP is deemed computationally expensive.

\begin{table}[tb]
\centering
\large
\resizebox{\columnwidth}{!}{%
\begin{tabular}{lcccccccc}
\toprule
\textbf{Algorithm} & $\lambda$ &
\multicolumn{2}{c}{\textbf{Known VA}} &
\multicolumn{2}{c}{\textbf{Learning + Mean VA}} &
\multicolumn{2}{c}{\textbf{Learning + Sampled VA}} \\
& &
\multicolumn{2}{c}{\textbf{(\smartGS)}} &
\multicolumn{2}{c}{\textbf{(\smartmean)}} &
\multicolumn{2}{c}{\textbf{(\smartmsmpl)}} \\
\cmidrule(lr){3-4}\cmidrule(lr){5-6}\cmidrule(lr){7-8}
& & \textbf{Agreement} & \textbf{\overloadshort}
  & \textbf{Agreement} & \textbf{\overloadshort}
  & \textbf{Agreement} & \textbf{\overloadshort} \\
& & \textbf{Rate} &
  & \textbf{Rate} &
  & \textbf{Rate} & \\
\midrule
\leastload & -- & 0.406 & 25.13 & -- & -- & -- & -- \\
\alggrdstar & -- & 0.439 & 41.71 & 0.425 & 36.31 & -- & -- \\
\algThompsonstar & -- & -- & -- & -- & -- & 0.404 & 34.17 \\
\midrule
\multirow{4}{*}{\smart}
& 0.01 & \textcolor{darkgreen}{0.444} & \textcolor{red}{62.57}
       & \textcolor{red}{0.424} & \textcolor{red}{48.33}
       & \textcolor{darkgreen}{0.425} & \textcolor{red}{46.74} \\
& 0.05 & \textcolor{red}{0.438} & \textcolor{darkgreen}{40.54}
       & \textcolor{red}{0.418} & \textcolor{red}{34.92}
       & \textcolor{darkgreen}{0.423} & \textcolor{red}{36.05} \\
& 0.10 & \textcolor{red}{0.437} & \textcolor{darkgreen}{37.51}
       & \textcolor{red}{0.417} & \textcolor{darkgreen}{32.07}
       & \textcolor{darkgreen}{0.421} & \textcolor{darkgreen}{32.11} \\
& 0.50 & \textcolor{red}{0.436} & \textcolor{darkgreen}{36.83}
       & \textcolor{red}{0.417} & \textcolor{darkgreen}{31.00}
       & \textcolor{darkgreen}{0.420} & \textcolor{darkgreen}{30.90} \\
\midrule
\upbnd & -- & 0.460 & 132.15 & -- & -- & -- & -- \\
\bottomrule
\end{tabular}
}
\caption{Scenario 1 results with LP-Approximation.}
\label{tab:scenario1LP}
\end{table}

We repeat all of the experiments presented in the paper with \textbf{LP-Approximation} and provide the results in Tables  \ref{tab:scenario1LP}, \ref{tab:scenario2LP}, \ref{tab:fulldataGRNDLP} and \ref{tab:realDataLearnLP}. Notice that the results are very similar to the same simulations with the QP presented in Tables 2, 3, 4 and 5. Much of the analysis and discussions regarding the results remain valid. However, when LP-approximation is used, we notice slight lower resolution rates at lower $\lambda$s and  lower OCDM at higher $\lambda$s. Also, please note that `Least Load', `\alggrdstar', `TS*', and `Upper Bound' rows are verbatim copied from Tables 2, 3, 4 and 5 to \ref{tab:scenario1LP}, \ref{tab:scenario2LP}, \ref{tab:fulldataGRNDLP} and \ref{tab:realDataLearnLP} to help the reader compare \smart algorithms against baselines. Only the \smartGS, \smartmean and \smartmsmpl algorithms are impacted by using LP instead of the QP, and the corresponding  rows are re-simulated and updated.

\begin{table}[tb]
\centering
\large
\resizebox{\columnwidth}{!}{%
\begin{tabular}{lcccccccc}
\toprule
\textbf{Algorithm} & $\lambda$ &
\multicolumn{2}{c}{\textbf{Known VA}} &
\multicolumn{2}{c}{\textbf{Learning + Mean VA}} &
\multicolumn{2}{c}{\textbf{Learning + Sampled VA}} \\
& &
\multicolumn{2}{c}{\textbf{(\smartGS)}} &
\multicolumn{2}{c}{\textbf{(\smartmean)}} &
\multicolumn{2}{c}{\textbf{(\smartmsmpl)}} \\
\cmidrule(lr){3-4}\cmidrule(lr){5-6}\cmidrule(lr){7-8}
& & \textbf{Agreement} & \textbf{\overloadshort}
  & \textbf{Agreement} & \textbf{\overloadshort}
  & \textbf{Agreement} & \textbf{\overloadshort} \\
& & \textbf{Rate} &
  & \textbf{Rate} &
  & \textbf{Rate} & \\
\midrule
\leastload & -- & 0.380 & 4.13 & -- & -- & -- & -- \\

\alggrdstar & -- & 0.406 & 5.18 & 0.394 & 5.06 & -- & -- \\

TS* & -- & -- & -- & -- & -- & 0.392 & 4.82 \\
\midrule
\multirow{4}{*}{\smart}
& 0.01 & \textcolor{darkgreen}{0.447} & \textcolor{red}{70.09}
       & \textcolor{darkgreen}{0.413} & \textcolor{red}{47.16}
       & \textcolor{darkgreen}{0.407} & \textcolor{red}{36.99} \\
& 0.05 & \textcolor{darkgreen}{0.429} & \textcolor{red}{27.58}
       & \textcolor{darkgreen}{0.403} & \textcolor{red}{18.89}
       & \textcolor{darkgreen}{0.402} & \textcolor{red}{21.03} \\
& 0.10 & \textcolor{darkgreen}{0.416} & \textcolor{red}{11.49}
       & \textcolor{darkgreen}{0.398} & \textcolor{red}{10.38}
       & \textcolor{darkgreen}{0.395} & \textcolor{red}{11.29} \\
& 0.50 & \textcolor{darkgreen}{0.413} & \textcolor{red}{5.53}
       & \textcolor{darkgreen}{0.397} & \textcolor{red}{5.79}
       & \textcolor{darkgreen}{0.394} & \textcolor{red}{5.42} \\
\midrule
\upbnd & -- & 0.460 & 132.15 & -- & -- & -- & -- \\
\bottomrule
\end{tabular}
}
\caption{Scenario 2 results with LP-Approximation.}
\label{tab:scenario2LP}
\end{table}

\begin{table}[tb]
\centering
\large
\resizebox{0.8\columnwidth}{!}{%
\begin{tabular}{lcccc}
\toprule
\textbf{Algorithm} & $\lambda$ &
\textbf{Agreement} & \textbf{\overloadshort} &
\textbf{Caseload} \\
& & \textbf{Rate} & & \textbf{Gini Index} \\
\midrule
\leastload
& -- & 0.467 & 0.10 & 26.1\% \\

\alggrdstar
& -- & 0.569 & 0.32 & 74.6\% \\
\midrule
\multirow{7}{*}{\smartGS}
& 0.01 & \textcolor{darkgreen}{0.615}
       & \textcolor{red}{49.37}
       & \textcolor{lightred}{87.1\%} \\
& 0.02 & \textcolor{darkgreen}{0.600}
       & \textcolor{red}{24.45}
       & \textcolor{lightred}{82.1\%} \\
& 0.03 & \textcolor{darkgreen}{0.589}
       & \textcolor{red}{12.10}
       & \textcolor{lightred}{78.7\%} \\
& 0.04 & \textcolor{darkgreen}{0.581}
       & \textcolor{red}{6.05}
       & \textcolor{darkred}{76.6\%} \\
& 0.05 & \textcolor{darkgreen}{0.578}
       & \textcolor{red}{3.16}
       & \textcolor{darkred}{75.4\%} \\
& 0.10 & \textcolor{darkgreen}{0.571}
       & \textcolor{red}{0.39}
       & \textcolor{lightgreen}{74.0\%} \\
& 0.50 & \textcolor{darkgreen}{0.570}
       & \textcolor{darkgreen}{0.31}
       & \textcolor{lightgreen}{73.9\%} \\
\midrule
\upbnd
& -- & 0.659 & 122.01 & 96.5\% \\
\bottomrule
\end{tabular}
}
\caption{Simulations with true VA information available to the allocation algorithms on data collected from the judiciary with LP-Approximation.}
\label{tab:fulldataGRNDLP}
\end{table}

\begin{table}[t]
\centering
\large
\resizebox{\columnwidth}{!}{%
\begin{tabular}{lcccccccc}
\toprule
\textbf{Algorithm} & $\lambda$ &
\multicolumn{3}{c}{\textbf{Initial belief set to $\mathcal{N}(0,\sigma_{\mu}^2)$}} &
\multicolumn{3}{c}{\textbf{Initial belief calculated from data}} \\
\cmidrule(lr){3-5}\cmidrule(lr){6-8}
& & \textbf{Agreement} & \textbf{\overloadshort} & \textbf{Caseload}
  & \textbf{Agreement} & \textbf{\overloadshort} & \textbf{Caseload} \\
& & \textbf{Rate} & & \textbf{Gini Index}
  & \textbf{Rate} & & \textbf{Gini Index} \\
\midrule
\alggrdstar
& -- & 0.489 & 0.44 & 68.4\%
     & 0.473 & 0.38 & 63.1\% \\
\midrule
\multirow{4}{*}{\smartmean}
& 0.01 & \textcolor{darkgreen}{0.514}
       & \textcolor{red}{70.00}
       & \textcolor{lightred}{87.0\%}
       & \textcolor{red}{0.444}
       & \textcolor{red}{30.29}
       & \textcolor{lightred}{71.5\%} \\
& 0.05 & \textcolor{darkgreen}{0.496}
       & \textcolor{red}{10.32}
       & \textcolor{darkred}{72.0\%}
       & \textcolor{red}{0.465}
       & \textcolor{red}{0.70}
       & \textcolor{lightgreen}{61.2\%} \\
& 0.10 & \textcolor{darkgreen}{0.490}
       & \textcolor{red}{1.22}
       & \textcolor{lightgreen}{68.3\%}
       & \textcolor{red}{0.471}
       & \textcolor{red}{0.38}
       & \textcolor{lightgreen}{61.5\%} \\
& 0.50 & \textcolor{darkgreen}{0.490}
       & \textcolor{darkgreen}{0.42}
       & \textcolor{lightgreen}{68.0\%}
       & \textcolor{red}{0.471}
       & \textcolor{darkgreen}{0.37}
       & \textcolor{lightgreen}{61.9\%} \\
\midrule
\algThompsonstar
& -- & 0.484 & 0.34 & 53.7\%
     & 0.489 & 0.31 & 50.8\% \\
\midrule
\multirow{4}{*}{\smartmsmpl}
& 0.01 & \textcolor{darkgreen}{0.496}
       & \textcolor{red}{21.69}
       & \textcolor{darkdarkred}{57.5\%}
       & \textcolor{darkgreen}{0.492}
       & \textcolor{red}{6.14}
       & \textcolor{lightgreen}{49.4\%} \\
& 0.05 & \textcolor{darkgreen}{0.487}
       & \textcolor{red}{3.88}
       & \textcolor{darkgreen2}{50.8\%}
       & \textcolor{red}{0.489}
       & \textcolor{red}{0.82}
       & \textcolor{darkgreen}{47.5\%} \\
& 0.10 & \textcolor{darkgreen}{0.485}
       & \textcolor{red}{0.74}
       & \textcolor{darkgreen2}{49.7\%}
       & \textcolor{darkgreen}{0.490}
       & \textcolor{darkgreen}{0.28}
       & \textcolor{darkgreen}{47.7\%} \\
& 0.50 & \textcolor{red}{0.483}
       & \textcolor{darkgreen}{0.28}
       & \textcolor{darkgreen2}{49.4\%}
       & \textcolor{darkgreen}{0.491}
       & \textcolor{darkgreen}{0.25}
       & \textcolor{darkgreen}{48.0\%} \\
\bottomrule
\end{tabular}
}
\caption{Simulations on data collected from the judiciary where VAs need to be learned during allocation with LP-Approximation.}
\label{tab:realDataLearnLP}
\end{table}

\subsection{Estimating uncertainty of mediator Value-Addeds}
In this section, we clarify the mathematics underlying the uncertainty estimation for mediator VAs.  We denote the outcome of case $i$ assigned to mediator $j$ using the random variable $Y_{ij}$. $Y_{ij}=1$ when the case is resolved successfully, and $0$ otherwise. Also, let us recall the probability of successful resolution in this case would be given by $p_i+\mu_j$. In other words:

\resizebox{0.6\linewidth}{!}{$
Pr[Y_{ij} \mid \mu_i] =
\begin{cases}
\mu_j+p_i,
& \text{If } Y_{ij}  = 1, \\[10pt]
1-(\mu_j+p_i),
& \text{If } Y_{ij}  = 0 
\end{cases}
$}

\begin{align*}
f(\mu_j \mid Y_{ij}=1)
&= \frac{\Pr(Y_{ij}=1 \mid \mu_j)\, f(\mu_j)}
{\int_{-\infty}^{\infty}\Pr(Y_{ij}=1 \mid \mu_j)\, f(\mu_j)\, d\mu_j}\\
&= \frac{(\mu_j+p_i)\,\frac{1}{\sqrt{2\pi}\sigma_\mu}
\exp\!\left(-\frac{\mu_j^2}{2\sigma_\mu^2}\right)}
{\int_{-\infty}^{\infty}(\mu_j+p_i)\,\frac{1}{\sqrt{2\pi}\sigma_\mu}
\exp\!\left(-\frac{\mu_j^2}{2\sigma_\mu^2}\right)\, d\mu_j}\\
&\propto (\mu_j+p_i)\exp\!\left(-\frac{\mu_j^2}{2\sigma_\mu^2}\right).
\end{align*}
\begin{figure*}[h]
  \centering
  \includegraphics[width=0.8\textwidth]{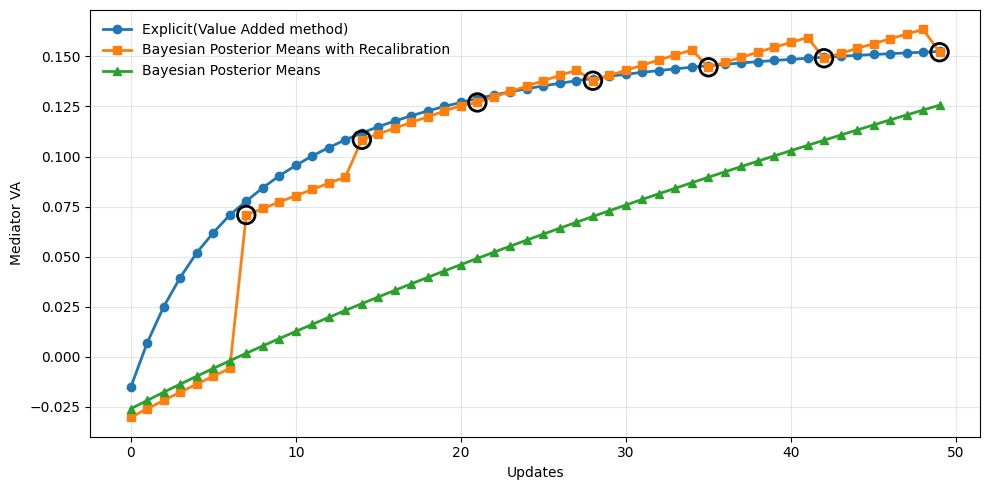}
  \caption{Trajectory of a single mediator’s value-added (VA) estimate over 50 sequential successful case-outcome updates under three methods. The Bayesian posterior mean without recalibration (green) drifts away from the explicit VA estimate (blue), while periodic recalibration applied every seventh update (orange; circled points) limits this drift and keeps posterior estimates closely aligned with the explicit VA.}
  \label{fig:recalibration}
\end{figure*}
Computing the mediator VAs explicitly using methods outlined in section 3.2 provides us a global mediator level standard deviation $\sigma_\mu$. We assume a prior $\mathcal{N}(0,\sigma_\mu)$ on the mediator VAs. With that assumption, we get:
\begin{align*}
    f(\mu_j) = \frac{1}{\sqrt{2\pi}\sigma_\mu} e^{(-\frac{\mu_j^2}{2\sigma_\mu^2}) }
\end{align*}

Now, if we make this assignment and observe the outcome $Y_{ij}$, then a Bayesian posterior can be used to compute:
\begin{align}
    f(\mu_i\mid Y_{ij}) = \frac{Pr[Y_{ij}\mid \mu_j] f(\mu_j)}{\int_{-\infty}^{\infty}Pr[Y_{ij}\mid \mu_j] f(\mu_j)d\mu_j} \label{eq:bayes_update}
\end{align}

We now simplify equation \ref{eq:bayes_update} on a case by case basis.

When $Y_{ij} = 1$:

Similarly, when $Y_{ij} = 0$, we can simplify to:

\begin{align*}
    f(\mu_j\mid Y_{ij}=0) &\propto (1-(\mu_j+p_i))e^{(-\frac{\mu_j^2}{2\sigma_\mu^2}) }
\end{align*}

Combining both, we can write: 
\begin{align}
    f(\mu_j\mid Y_{ij}) &\propto (\mu_j+p_i)^{Y_{ij}}(1-(\mu_j+p_i))^{1-Y_{ij}}exp{(-\frac{\mu_j^2}{2\sigma_\mu^2}) } \label{eq:finalupdate}
\end{align}

Now, for the next case $i'$ being assigned to the same mediator $j$, we can use the posterior in Eq \ref{eq:finalupdate} as a prior and compute the posterior $f(\mu_j\mid Y_{ij},Y_{i'j})$. However, we have found using $f(\mu_j\mid Y_{ij})$ directly as a prior makes the analysis complicated. As a remedy, we use gaussian moment matching to define the posterior $f(\mu_j\mid Y_{ij})$ to be  $\mathcal N(\mu_j \mid Y_{ij}, (\sigma_j \mid Y_{ij})^2)$ where $\mu_j \mid Y_{ij}=E[f(\mu_j\mid Y_{ij})]$ and $(\sigma_j \mid Y_{ij})^2 = Var[f(\mu_j\mid Y_{ij})]$. Using this method, we use gaussian priors to compute posteriors as successive case outcomes become available.

Finally, the standard deviations associated with these posteriors are used as the uncertainty associated with  mediator $j$'s VA.

\subsection{Efficient VA updates and Re-calibration of the posteriors }
Ideally during the simulation, mediator VAs should be calculated explicitly using the VA method outlined in section 3.2. This computation needs to be done every time an assigned case reaches conclusion. The explicit VA calculation method requires processing all case outcomes allocated to all mediator until that point. As a result, the explicit method is computationally prohibitive to repeat for every case allocation. 

Computing the VA uncertainties using Bayesian posterior method can provide us a computationally inexpensive estimate of the mediator VAs in the form of posterior means. However,  Gaussian moment matching introduces inaccuracies in the posteriors which can add up to be significant. Therefore, these estimates are rather unreliable as a substitute for explicit VA computations.

We have found the hybrid approach of recalibrating the posterior means by resetting them to VAs computed explicitly to be a good trade off. In the simulation, we start off with explicit VA estimations. When cases are resolved, we perform the immediate VA updates using posterior means. And every 7 days, we explicitly re-compute mediator VAs using the Value Added method with updated case histories, and then recalibrate the Bayesian posteriors. 

To illustrate the accuracy gains from recalibration, we track the estimated VA of a single mediator over 50 sequential case-outcome updates under each of the three methods. In this experiment, all 50 case outcomes are successful, and recalibration is performed every seventh update. Figure \ref{fig:recalibration} shows the resulting trajectories. Without recalibration, the Bayesian posterior mean (green) can drift substantially away from the explicit VA estimate (blue). By contrast, periodic recalibration (orange) constrains this drift, keeping the posterior estimates within a tolerable range of the explicit VAs.

\end{document}